\documentclass[twocolumn,english,prl,aps,showpacs,superscriptaddress]{revtex4-1}
\usepackage[latin9]{inputenc}
\usepackage{verbatim}
\usepackage{amsmath}
\usepackage{amssymb}
\usepackage{graphicx}
\usepackage{esint}

\makeatletter

\newcommand{\R} {R_{B1g}(\Omega)}

\usepackage{bm}
\usepackage{xcolor}
\usepackage{subfigure}
\usepackage{array}

\@ifundefined{textcolor}{}{%
 \definecolor{BLACK}{gray}{0}
 \definecolor{WHITE}{gray}{1}
 \definecolor{RED}{rgb}{1,0,0}
 \definecolor{GREEN}{rgb}{0,1,0}
 \definecolor{BLUE}{rgb}{0,0,1}
 \definecolor{CYAN}{cmyk}{1,0,0,0}
 \definecolor{MAGENTA}{cmyk}{0,1,0,0}
 \definecolor{YELLOW}{cmyk}{0,0,1,0}
}

\def\Gr{\mathcal G}

\usepackage{bm}

\newcommand{\be}{\begin{equation}}
\newcommand{\ee}{\end{equation}}
\newcommand{\bea}{\begin{eqnarray}}
\newcommand{\eea}{\end{eqnarray}}
\newcommand{\beq}{\begin{equation}}
\newcommand{\eeq}{\end{equation}}
\def\lb{\label}

\newcommand{\bk}{\mathbf{k}}
\newcommand{\bq}{\mathbf{q}}

\newcolumntype{x}[1]{>{\centering\arraybackslash}p{#1}}

\usepackage{babel}

\makeatother

\def\lb{\label}
\def\pref#1{(\ref{#1})}

\newcount\bozza \bozza=0
\ifnum\bozza=1
\newdimen\shift \shift=-2truecm
\def\lb#1{%
{\label{#1}\rlap{\kern\shift{$\scriptstyle#1$}}}}
\else\def\lb#1{\label{#1}} \fi

\begin{document}

\title{Raman response in the nematic phase of FeSe}

\author{Mattia Udina}

\selectlanguage{english}

\affiliation{Department of Physics, ``Sapienza'' University of Rome, P.le
A. Moro 5, 00185 Rome, Italy}
\affiliation{Institute for Complex Systems (ISC-CNR), UOS Sapienza, P.le A. Moro 5, 00185 Rome, Italy}

\author{Marco Grilli}

\affiliation{Department of Physics, ``Sapienza'' University of Rome, P.le
A. Moro 5, 00185 Rome, Italy}

\affiliation{Institute for Complex Systems (ISC-CNR), UOS Sapienza, P.le A. Moro 5, 00185 Rome, Italy}

\author{Lara Benfatto}

\affiliation{Department of Physics, ``Sapienza'' University of Rome, P.le
A. Moro 5, 00185 Rome, Italy}
\affiliation{Institute for Complex Systems (ISC-CNR), UOS Sapienza, P.le A. Moro 5, 00185 Rome, Italy}

\author{Andrey V. Chubukov}
\affiliation{School of Physics and Astronomy, University of Minnesota, Minneapolis,
MN 55455, USA}

\date{\today}

\begin{abstract}
Raman experiments on bulk FeSe revealed
that the low-frequency part of  $B_{1g}$ Raman response $\R$,  which probes nematic fluctuations,
rapidly decreases below the nematic transition at  $T_n \sim 85$K.
Such behavior is usually associated with the gap opening and at a first glance is inconsistent with the fact that FeSe
 remains a metal below $T_n$,  with sizable   hole and electron pockets. We argue that
  the drop of $\R$  in a nematic metal
 comes about because the nematic order drastically changes the orbital content of the pockets and makes them nearly mono-orbital.
In this situation
  $B_{1g}$ Raman response
   gets reduced by the same
   vertex corrections
    that enforce charge conservation.
   The reduction holds at low frequencies and gives rise to gap-like behavior of $\R$,
   in full agreement with the experimental data.
\end{abstract}
\maketitle

\emph{Introduction.}
Bulk Fe-chalcogenide FeSe has recently attracted a lot of attention due to  a peculiar property, not seen in other Fe-based superconductors --
the emergence of a nematic order (a spontaneous breaking of $C_4$ lattice symmetry down to $C_2$) below
$T_n \sim 85$ K, without antiferromagnetism nearby~\cite{Bohmer_review17,ColdeaWatsonreview18,Fernandesreview17,GallaisReview16}.
 Extensive  STM and ARPES studies~\cite{Suzuki15, Watsonprb15,Watsonprb16,Fanfarilloprb16,WatsonNJP17,Watson_prb17,Davis2017,BorisenkoScienRep17,Borisenkoprb18,Watson_prb18,Hashimoto-ShinNatComm18,Liu_PRX2018,Yiarxiv19,Kim_arxiv19}
 revealed a sizeable deformation of hole and electron  Fermi surfaces (FSs) below $T_n$, which are well reproduced by introducing an electronic orbital splitting
 \cite{Fanfarilloprb16,Davis2017,Hirschfeld2017,Chubukov_prl18,Fanfarillo_qm18}.

A remarkable indication of the electronic nature of the nematic transition in FeSe has been provided by Raman measurements
~\cite{MassatGallaispnas16,Thorsmolle2016,Blumbergarxiv17,hackl19}.  In a metal, the Raman response probes density-like fluctuations
   at a finite frequency $\Omega$  and vanishing momenta $q$,
  modulated by a form factor, which depends on relative polarizations of the incoming and outgoing light and transforms according to the point-group representation of the crystal
  ~\cite{devereaux_review}.
   When the Raman factor
   reduces to a constant as, e.g., in a fully symmetric channel in a single-band system,
the Raman response is proportional to
 the density correlator
 and vanishes
 at $q=0$ and finite $\Omega$,  because
  fermionic density is a conserved quantity~\cite{deveraux_review,kleinmgb2,GallaisReview16,yamase,kontani,Klein18,Klein18a,cea_prb16,maiti_prb17}.
The response that probes electronic nematic correlations is in the non-symmetric  $B_{1g}$ scattering channel~\cite{Note1},
 and a generic belief is that it is finite in a metal, because no conservation law applies~\cite{GallaisReview16,yamase,kontani,Klein18,Klein18a}.
 Above the nematic transition, the measured profile of the Raman intensity $\R$
    at small frequencies
    is well approximated by
    $R_{B_{1g}} (\Omega) \propto  \mbox{Im} [\chi_{B_{1g}}(\Omega)/(1-U \chi_{B_{1g}} (\Omega)]$, where
$U$ is the attractive interaction in the $d-$wave particle-hole channel, and
the $B_{1g}$ Raman susceptibility  has a conventional  relaxation   form~\cite{GallaisReview16,Karahasanovic2015,yamase}
$\chi_{B_{1g}}(\Omega) = \chi_{B_{1g}}(q=0,\Omega) \propto 2i \gamma/(\Omega + 2i\gamma)$,
where the scattering rate $\gamma$ is either due to impurities~\cite{GallaisReview16} or to electron-electron interaction~\cite{Klein18,Klein18a}.
This yields

\beq
    \R \propto \frac{\Omega \gamma}{\Omega^2 + 4\gamma^2 (1 - U/U_{cr})^2},
    \lb{new_2}
    \eeq
    where
      $U_{cr}$ is the value of $U$ at which the system develops a nematic order.
    As $U$ approaches $U_{cr}$, the peak width narrows as $ 1 - U/U_{cr}$ and the
   intensity of the peak increases as $1/(1 - U/U_{cr})$.  Both results agree with the data, which show
 that the peak in $\R$ narrows and moves to a lower energy as  $T$ approaches $T_n$.

  However, this agreement holds only at $T$ above the nematic transition.
  Below $T_n$, the data show that $\R$ rapidly drops
   at $\Omega \lesssim 200$ cm$^{-1}$.  Such behavior is expected  when
   quasiparticles  acquire a finite gap (e.g., when superconductivity develops\cite{cea_prb16,maiti_prb17}). However, FeSe remains a metal in the nematic phase,
   with deformed, but still sizable hole and electron pockets. We show below  that
       the Fermi-surface deformation
       has little effect on
       the behavior of  $\R$ at small frequencies.
        We argue that the origin of
the gap-like behavior of $\R$
 is the change of the
 orbital composition
   of the pockets
   below $T_n$.

  The outline of our reasoning is as follows.  In the tetragonal phase,
 FeSe  has
  two  nearly circular hole pockets
   at the zone center $\Gamma$  and two electron pockets at $M = (\pi,\pi)$ (in the
 2Fe Brillouin-zone notation), split by spin-orbit coupling into inner and outer pockets~\cite{comment1}.
  The hole and the inner electron pocket are
  constructed out of
  $d_{xz}$ and $d_{yz}$ orbitals, and the outer electron pocket has predominantly  $d_{xy}$ character.
   The $B_{1g}$  Raman vertex is
   $\Gamma_{B_{1g}} = d^\dagger_{xz} d_{xz} - d^\dagger_{yz} d_{yz}$, where $d^\dagger, d$ are creation and
   annihilation operators for the corresponding orbitals
   \cite{GallaisReview16,Thorsmolle2016,Hinojosa2016}.
 In the band basis,
 $\Gamma_{B_{1g}}$
 is then highly sensitive to the orbital composition of the pockets.
 In the tetragonal phase $\Gamma_{B_{1g}}$ has  pure $d$-wave symmetry,  e.g., near
    the outer
     hole pocket
  it is $d^\dagger_h d_h \cos{2\theta}$, where $d^\dagger_h, d_h$ are band operators and $\theta$ is the angle along the pocket.
    The  nematic order parameter $\Delta$ breaks  $C_4$ symmetry between the
   orbital occupations and induces an
   additional term  $\Delta_h (d^\dagger_{xz} d_{xz} - d^\dagger_{yz} d_{yz})$ in the Hamiltonian~\cite{comment2}.
    When this order develops, it not only  elongates  hole and inner electron pockets in the directions set by the signs of $\Delta$ at $\Gamma$
    and at $M$ (Refs. \cite{kontani_1,Fanfarilloprb16,Davis2017,Hirschfeld2017,Chubukov_prl18,Fanfarillo_qm18}), but also changes the orbital content of the pockets.
      In FeSe,
     the pockets are small,
   and the orbital content changes quite drastically.       Calculations~\cite{Fanfarilloprb16,Hirschfeld2017,Chubukov_prl18,Fanfarillo_qm18}
       and polarized ARPES data~\cite{Watson_prb18,Liu_PRX2018}  show that deep inside the nematic phase the
       outer
        hole pocket becomes predominantly $d_{xz}$
         electron pocket
          becomes
           predominantly
           $d_{yz}$, or vice versa.
  This in turn affects
   $\Gamma_{B_{1g}}$, which develops
    an angle-independent
     component, proportional to the nematic order parameter.
     When the Fermi pockets become nearly mono-orbital,  this component becomes the dominant one and  $B_{1g}$ Raman
  susceptibility  becomes almost identical to
   the susceptibility in the density channel.
    The latter, we remind, vanishes
     at $q=0$ and finite $\Omega$
  by charge conservation~\cite{Klein18,Klein18a,cea_prb16,maiti_prb17}.
Accordingly,
the $B_{1g}$ Raman intensity
 should also get strongly reduced.
   This holds at $\Omega \leq
   2 -3
   \Delta$.
   At larger $\Omega$ the electronic excitations recover the same orbital character of the tetragonal phase
    and $\R$ rapidly increases.

 At the computational level, the reduction of $\R$ is associated with the effect of vertex corrections, which must be included along with the fermionic damping $i \gamma$, once the Raman vertex
  acquires an $s$-wave component~\cite{aleiner}.
  The damping rate itself does not change much by nematic order, again because the system remains a metal.
  The exact form of the reduced $\R$ at small frequencies depends on the details of fermionic dispersion, as we will show below. However, the reduction of the Raman response in the $B_{1g}$ channel in the nematic phase due to the change of the orbital composition of the pockets is a rather general and robust result.
Below we discuss in detail the contribution to $\R$ from the hole
pockets
  at $\Gamma$. The
  contribution to $\R$  from the
 electron pockets
  is analyzed
  along the same lines.

\begin{figure}[t]
\begin{center}
\includegraphics[scale=0.5,clip=true]{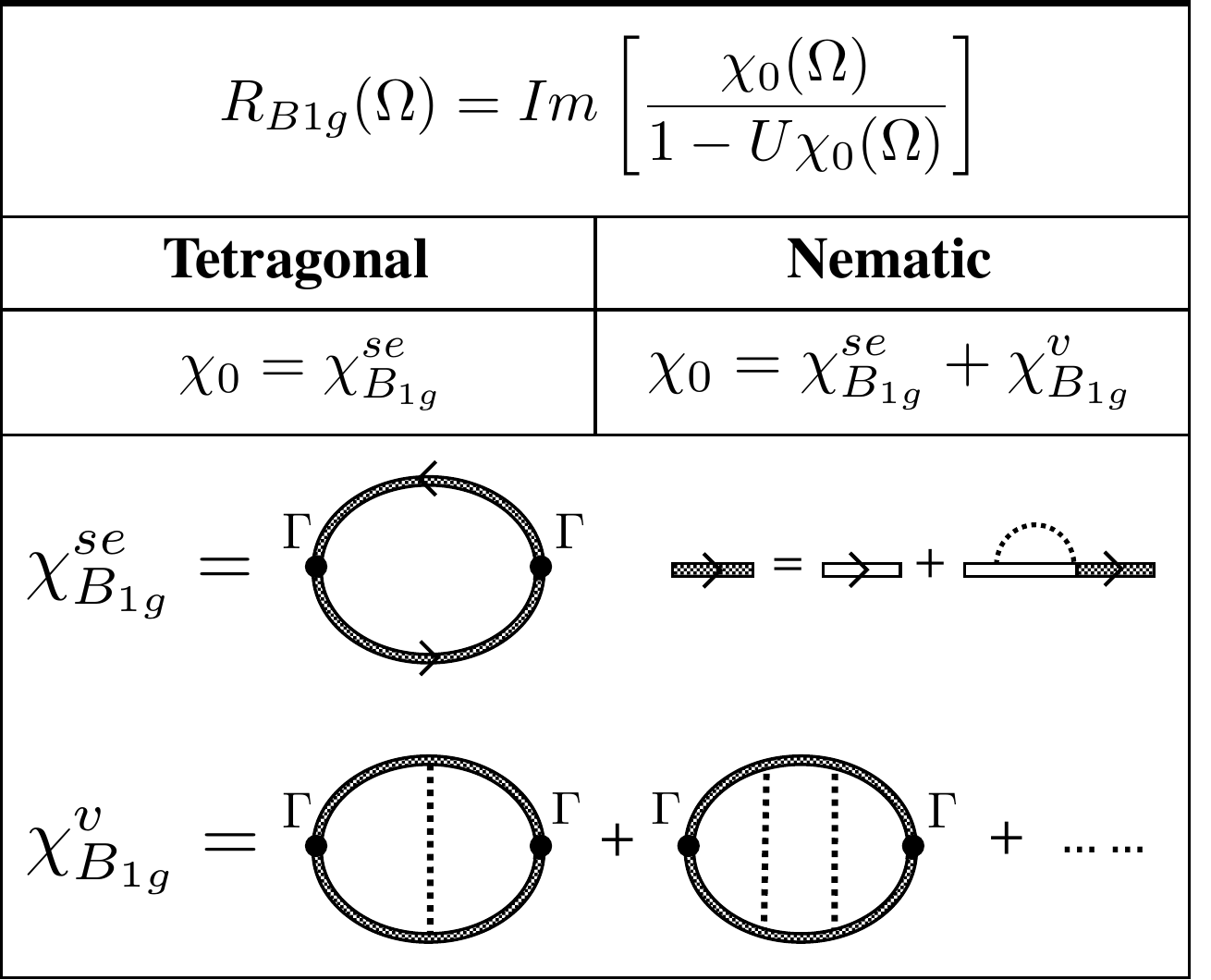}
\caption{Diagrammatic representation of the Raman susceptibility for the case when impurity scattering rate has only $s-$wave component. In the tetragonal phase, Raman vertex has pure $d-$wave form, and
 vertex corrections due to impurity scattering vanish. In the nematic phase, the vertex has both $d$-wave and $s$-wave components.
 The $s-$wave part of the Raman susceptibility is eliminated by vertex corrections.}
\label{fig3new}
\end{center}
\vspace{-0.9cm}
\end{figure}

 \emph{Orbital composition of the hole pocket.}~~~ The low-energy states near the hole pockets
    are described by the effective Hamiltonian~\cite{Vafekprb13}
 \bea
 &&H = \label{eq:new1} \\
 &&\sum_{\mathbf{k}, \sigma} \Psi_{\mathbf{k}, \sigma}^{ \dag} h_0 \tau_0 + h_1 \tau_1 + \eta \tau_2 + (h_3+\Delta_{h})\tau_3 \Psi_{\mathbf{k}, \sigma}, \nonumber
   \eea
  where  $\Psi \equiv (d_{xz}, d_{yz})$ is a spinor in the orbital space, $h_0 (\bk) = \epsilon_0- k^2/(2m), ~h_1 (\bk) =  2c k_x k_y$, and $h_3 (\bk) = - b (k^2_x-k^2_y)$ are
 hopping integrals in momentum space ($\epsilon_0$ includes the static self-energy that accounts for
   the shrinking of
     the hole pockets
     ~~\cite{Fanfarilloprb16,Fanfarillo2018,Fanfarillo_qm18}),  $\Delta_h = \Delta_h (T)$ is the magnitude of the nematic order, which we set to be positive, and $\eta$ is the spin-orbit
  (${\bf L} {\bf S}$) coupling.  The nematic order can originate from
    a $d-$wave Pomeranchuk instability~\cite{Chubukovprx16,kontani,kontani_1} or from composite spin fluctuations~\cite{RMF14,Fanfarilloprb16,Fanfarillo2018}.
     For our purposes,    the microscopic origin of the nematic order parameter is not important, and
     in the following we will just compare the Raman response in the presence and in the absence of $\Delta_{h}$.
  To simplify the presentation, in analytical formulas below we  set $b=c >0$, in which case the pocket in the tetragonal
  phase is circular. For numerical calculations we choose the hopping parameters to best match the data on FeSe.

The transformation from the orbital to the band basis  is described by the unitary matrix
 with components $u^2_\bk = \cos^2{{\bar \theta}_\bk}, |v_\bk|^2 =
\sin^2{{\bar \theta}_\bk}$,
 where
 \bea
 \cos{2 {\bar \theta}_\bk}  &=& \frac{b k^2 \cos{2 \theta} - \Delta_h}{\sqrt{b^2 k^4 + \Delta^2_h -2 b k^2 \Delta_h \cos{2 \theta} + \eta^2}}
\lb{new_1}
 \eea
 and
  $\theta$ is the angle along the hole pocket.  At $\eta = \Delta_h =0$, two hole bands necessary cross the Fermi level and form the inner and the outer hole pockets.
 When $\eta$ or $\Delta_h$ (or both) are non-zero, the inner pocket gets smaller and may sink below the Fermi level.
 In FeSe, only the outer hole pocket has been detected in the nematic state, and
  in analytical treatment we focus on the corresponding band,
 i.e., approximate the Hamiltonian in the band basis as
  $H = \sum_\bk E_\bk d^\dagger_{h,\bk} d_{h,\bk}$ with $E_\bk = \epsilon_0 - k^2/(2m) + \sqrt{b^2 k^4 + \Delta^2_h -2 b k^2 \Delta_h \cos{2 \theta+\eta^2}}$.
 The weight of $d_{xz}$ and $d_{yz}$ orbitals
  is given by $|v_\bk|^2 = (1- \cos{2 {\bar \theta}_\bk})/2$ and $u^2_\bk = (1+ \cos{2 {\bar \theta}_\bk})/2$, respectively. In the tetragonal phase
 $\cos{2 {\bar \theta}_\bk} \propto  \cos{2 \theta}$,
   and $d_{xz}$ and $d_{yz}$ orbitals equally contribute to  band excitations.
In the nematic phase,  the weight depends on the ratio $\lambda = \Delta_h/(bk^2)$. When $\lambda$ is small, the two orbitals still contribute
 nearly equally. However, when $\lambda \gg 1$,  $\cos{2 {\bar \theta}_k} \approx -1$, independent on $\theta$. In this limit,  the hole pocket is made almost entirely of $d_{xz}$ orbitals.
  In FeSe, its value at the Fermi level, $\lambda_F \sim \Delta_h/(b k^2_F)$, is larger than in other Fe-based superconductors because $E_F \leq 10 meV$ is smaller.  Recent polarized ARPES study did find~\cite{Watson_prb18,Liu_PRX2018} that
   the weight of the  $d_{xz}$ is over $80\%$
   along the hole pocket.

\begin{figure}[t]
\begin{center}
\includegraphics[scale=0.5,clip=true]{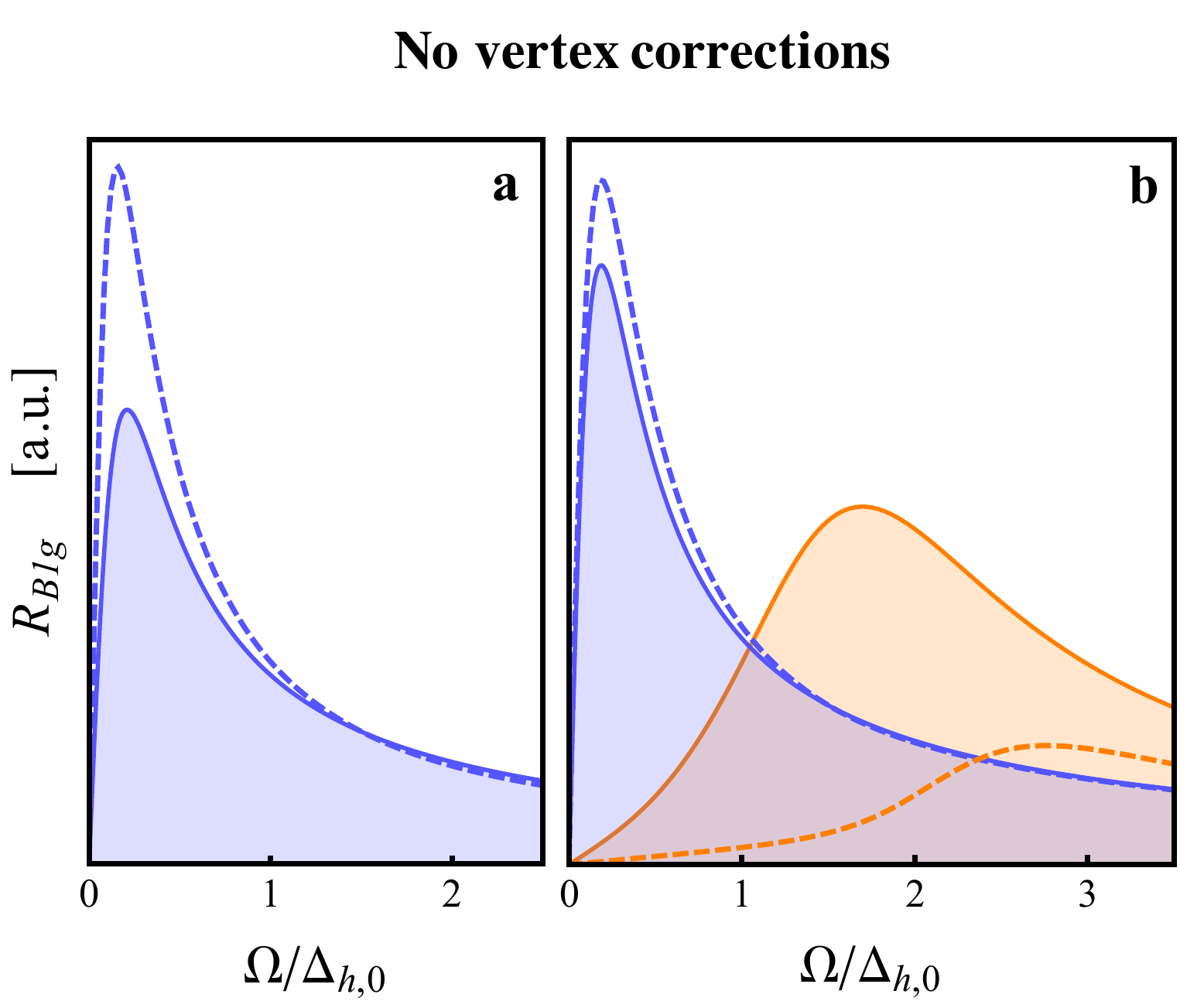}
\caption{Raman intensity in $B_{1g}$ channel, obtained neglecting vertex corrections. Solid/dashed lines denote the result in the tetragonal/nematic state. (a)
Intraband response
 for an outer hole pocket. (b) Combined response from inner and outer hole pockets. Blue lines -- the sum of intraband responses; orange line -- interband response.
The intraband response primarily comes  from the outer hole pocket and actually increases in the nematic state. The interband response is smaller and it gets reduced in the nematic state because the inner hole pockets sinks below the Fermi level.}
\label{fig1new}
\end{center}
\vspace{-0.9cm}
\end{figure}
\emph{The Raman response in the nematic phase.}~~~ The change of the orbital composition of the excitations has a profound
effect on the Raman response in the nematic phase. The $B_{1g}$ Raman vertex in the band basis is  expressed as\cite{suppl} $d^\dagger_{h,\bk} d_{h,\bk} \cos{2{\bar \theta}_\bk}$, and  the Raman intensity is
$R_{B_{1g}} (\Omega) \propto  \mbox{Im} [\chi_{B_{1g}}(\Omega)/(1-U \chi_{B_{1g}} (\Omega)]$, where
$\chi_{B_{1g}}(\Omega) = \chi_{B_{1g}}(q=0,\Omega)$ is the
 fully dressed particle-hole susceptibility  with Raman side vertices $\cos{2{\bar \theta}_\bk}$,
 and $U$ is the attractive interaction in the $d-$wave particle-hole channel.
  The diagrams for the Raman susceptibility are presented  in Fig. \ref{fig3new} under the simplifying assumption that the
   scattering rate
    $\gamma$ is angle-independent.
   The dressed bubble is $\chi_{B_{1g}} (\Omega) = \chi^{se}_{B_{1g}} (\Omega) + \chi^v_{B_{1g}} (\Omega)$, where 
   $\chi^{se}_{B_{1g}} (\Omega) \propto 2 i \gamma/(\Omega + 2 i \gamma)$
   is the particle-hole bubble made of fermions with damping rate $\gamma$, and
  $\chi^{v}_{B_{1g}} (\Omega)$ accounts for vertex corrections.
In the tetragonal phase, $\cos{2{\bar \theta}_\bk} \propto \cos{2\theta}$.  One can easily verify that  vertex corrections due to impurity scattering vanish because $\oint d \theta  \cos{2{\bar \theta}_\bk}=0$.
 Then $\chi_{B_{1g}} (\Omega) = \chi^{se}_{B_{1g}} (\Omega)$.
    This yields $\R$ as in Eq. (\ref{new_2}). In the nematic phase,
$\R$ changes because the pocket becomes elliptical and because the orbital content of hole excitations changes. We verified (see Fig. \ref{fig1new}a) that the change of the pocket
shape from circular to elliptical only  weakly affects the Raman intensity.
 The second effect, however, strongly reduces the $B_{1g}$  Raman bubble. To understand this, we note from Eq.\ \pref{new_1} that in the nematic phase, in addition to the $d-$wave $\cos {2\theta}$ term, $\cos{2{\bar \theta}_\bk}$ acquires an angle-independent, $s-$wave  component, proportional to the nematic order parameter itself. Neglecting other contributions for simplicity,  we write
 $\cos{2{\bar \theta}_\bk} = \Gamma_{s,\bk} + \Gamma_{d,\bk} \cos{2 \theta}$.  For the $d-$wave component, vertex corrections are still irrelevant, but for $s-$wave
 component they are non-zero and have to be included.
 At small frequencies, we can approximate $\Gamma_{s,\bk}$ and $\Gamma_{d,\bk}$ by constants $\Gamma_{s,k_F} = \Gamma_s$ and $\Gamma_{d,k_F} = \Gamma_d$.
  Summing up ladder series of vertex corrections (see Fig. \ref{fig3new})
  we
   obtain\cite{suppl}
 \bea
 &&\chi^{se}_{B_{1g}} (\Omega) = N_F \frac{2i \gamma}{\Omega + 2 i \gamma} \left(\Gamma^2_s + \Gamma^2_d\right) \nonumber \\
 &&\chi^{v}_{B_{1g}} (\Omega) = - N_F \frac{2i \gamma }{\Omega + 2 i \gamma} \Gamma^2_s,
 \label{new_4a}
 \eea
where $N_F$ is the density of states at the Fermi level.
Adding up the two terms we find that the $s$-wave contribution cancels out, as expected since for a constant $\Gamma_s$ form factor the Raman
susceptibility is proportional to ordinary density (charge) susceptibility, and the latter vanishes at $q=0$ and finite $\Omega$ because of charge conservation.

The rest yields
\bea
&&\chi_{B_{1g}} (\Omega) = N_F \frac{2i \gamma }{\Omega + 2 i \gamma} \Gamma^2_d \nonumber \\
&&\R  \propto  \Gamma^2_d  \frac{\Omega \gamma}{\Omega^2 + 4 \gamma^2 (1-U \Gamma^2_d/U_{cr})^2}.
\label{new_4b}
\eea
The result can be straightforwardly
extended to the case when the damping rate has both $s-$wave and $d-$wave components, $\gamma_s$ and $\gamma_d$. In this situation,
 we find that that Eq. (\ref{new_4b}) holds, but
  $\gamma = \gamma_s - \gamma_d$.

\begin{figure}[t]
\begin{center}
\includegraphics[scale=0.5,clip=true]{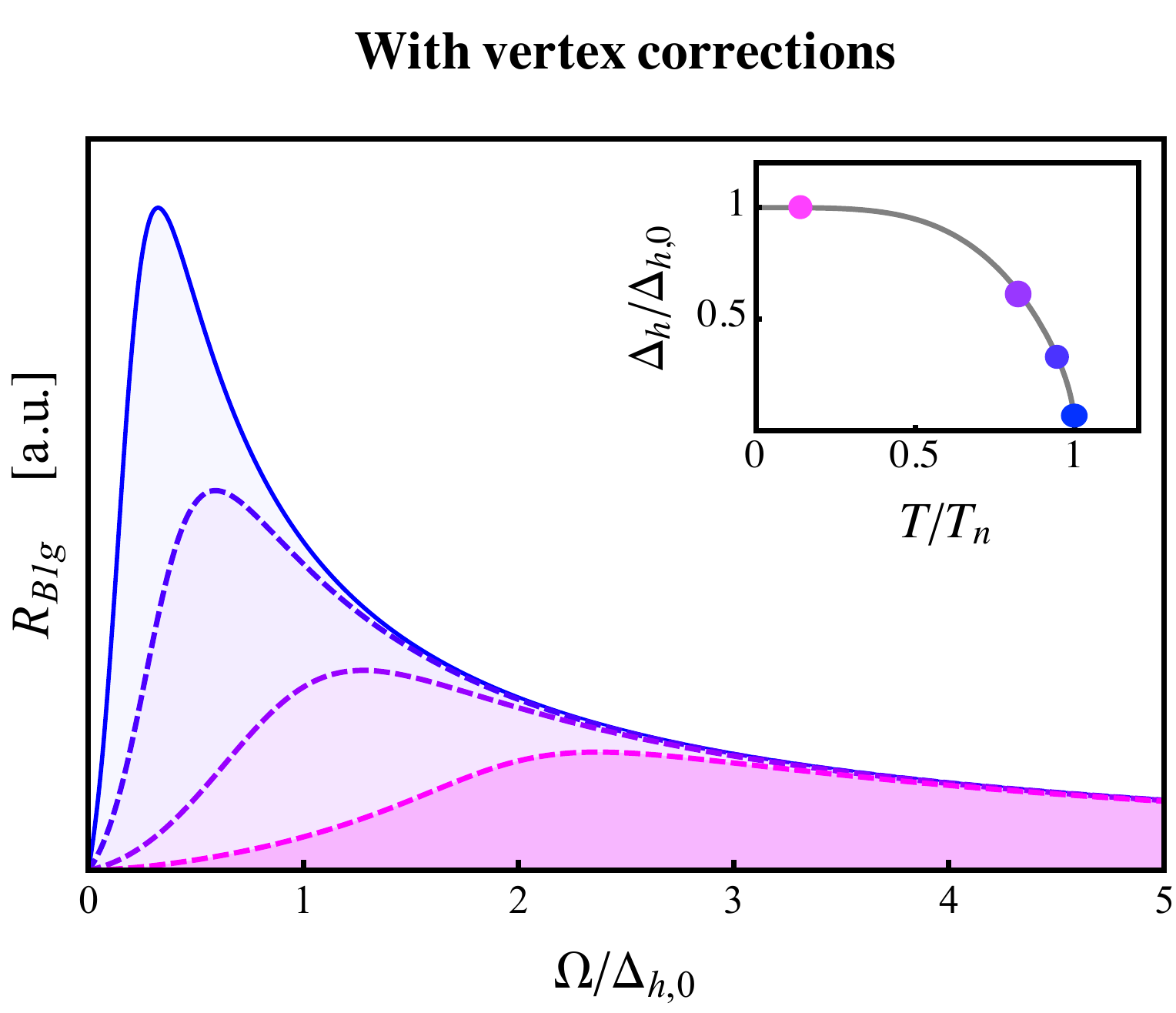}
\caption{Intraband Raman response
from the outer hole pocket,
 obtained by including vertex corrections with a frequency-dependent Raman vertex defined by Eq.\ \pref{gammas}-\pref{gammatot}.
Inset: temperature dependence of the nematic order parameter $\Delta_h(T)$ used for the calculations.
The Raman response recovers its form in the tetragonal phase at $\Omega \leq 3 \Delta_h$.
The addition of the inner hole pocket does not change this behavior.}
\label{fig2new}
\end{center}
\vspace{-0.9cm}
\end{figure}

We see that the full Raman $\R$ in the nematic phase retains the same functional form as in Eq. (\ref{new_2}), but the overall factor and $U$
are multiplied by $\Gamma^2_d$.   At large $\lambda_F$, when the Fermi hole pocket becomes almost entirely $d_{xz}$,  $\Gamma_s \approx -1$ and $\Gamma_d \sim 1/\lambda_F$ is small.
Then $\R$ is strongly reduced. The drop of $\R$ was phenomenologically attributed in ~\cite{blumberg_private} to the reduction in the damping rate.  In our theory, $\gamma$ does
not change, and the reduction is due to the small factor $\Gamma^2_d$.
The reduction holds only at small frequencies. At larger $\Omega$, typical fermionic momenta $k$ get larger,
$\lambda = \Delta_h/(bk^2)$ gets smaller, and
 $\Gamma_{d,\bk}$
increases. The Raman intensity
 $\R$ increases as well and eventually
 recovers its value in the tetragonal phase (see Fig. \ref{fig2new}).
Note in passing that long-range Coulomb interaction does not affect $\R$ by the same reasons as outlined in Refs. \cite{cea_prb16,maiti_prb17} -- the "mixed" bubble with $B_{1g}$
Raman vertex on one side and a constant ($s-$wave) vertex on the other, vanishes due to vertex corrections, even when both $\Gamma_s$ and $\Gamma_d$ are non-zero.

\emph{Numerical calculations.}~~~To set up a more realistic comparison with the experimental data we computed the Raman susceptibility $\chi_{B_{1g}}$ in the tetragonal and
nematic phases numericaly\cite{suppl}, using parameters appropriate to reproduce the band dispersion of FeSe, as measured by ARPES.  For $\chi^{se}_{B_{1g}}$, we found that the
 contribution to $\R$ from the outer hole pocket (left panel in Fig. \ref{fig1new})
   agrees well with the analytical
result, Eq. \pref{new_2}, over all the relevant frequency range, despite the fact that  Eq. \pref{new_2} has been  obtained by expanding near the Fermi surface\cite{suppl}.
 To obtain the full $\chi^{se}_{B_{1g}}$, one should also include interband contribution from particle-hole excitations between the two hole bands. We show the results in Fig. \ref{fig1new}b.
  Because  the inner hole pocket sinks below the Fermi level,  the inter-band contribution is small at low frequencies.
 The difference between $\R$ in the tetragonal and the nematic phases in this figure is due to nematicity-induced modification of the dispersions of the two bands.
  We see that the change of $\R$ is marginal:
    the intraband contribution gets slightly enhanced
     below $T_n$,
     and the interband one
      gets   reduced and
       shifts
        to larger frequencies.
The situation changes
 drastically when we include
  vertex corrections.  The most significant change is for the  intraband contribution, and we focus on this term.
     To obtain the intraband contribution to $\R$ in the full frequency range, we kept the dependence of $\cos 2\bar\theta_\bk$ on the momenta $\bk$, and did not restrict to $\bk \approx \bk_F$.
  Then $\Gamma_{s}$ and $\Gamma_{d}$ become functions of frequency, and Eq. (\ref{new_4a}) gets modified to
\bea
\lb{gammas}
\Gamma_s(\Omega)&=&\int \frac{d\theta kdk}{(2\pi)^2} \frac{\delta (\Omega-E_\bk)}{N(\Omega)}\cos 2\bar\theta_\bk,\\
\label{gammatot}
\Gamma_s^2(\Omega)+\Gamma_d^2(\Omega)&=&\int \frac{d\theta kdk}{(2\pi)^2} \frac{\delta (\Omega-E_\bk)}{N(\Omega)}(\cos 2\bar\theta_\bk)^2,
\eea
where $N(\Omega)=\int \frac{d\theta kdk}{(2\pi)^2} \delta (\Omega-E_\bk)$ is the frequency-dependent density of states.
The Raman response $\R$ is given by Eq.(\ref{new_4b}) with $\Gamma_d = \Gamma_d (\Omega)$.
  We recall that deep in the nematic phase, we have
    at small frequencies
    $\Gamma_s (\Omega)\approx -1$
and $\Gamma_d (\Omega) \approx 0$, This should lead to a near-complete suppression of $\R$ (see Eq.(\ref{new_4b})). As $\Omega$ increases, the relevant $k$ increase, and the $s$-wave component of
$\cos 2\bar\theta_\bk$ gets smaller. At large $k$ (high frequencies), $b k^2 \gg \Delta_h$, and for a generic $\theta$  one recovers $\cos 2\bar\theta_\bk \propto \cos 2\theta$. Then
 the Raman response should become the same as in the tetragonal phase.
 We show the result for $\R$  in  Fig. \ref{fig2new}.  We used the mean-field temperature dependence of $\Delta_h = \Delta_h (T)$, which is in good agreement with the temperature
 evolution of the band dispersion below $T_n$\cite{Fanfarilloprb16}. We see the behavior which we just outlined. Namely, at small $\Omega$  the intraband
 Raman response progressively develops a gap-like behavior as $T$ is lowered below $T_n$. At larger frequencies, $\R$ rapidly increases and becomes
  the same as in the tetragonal phase.
 This behavior is  in full agreement with the experimental data in~\cite{MassatGallaispnas16,Blumbergarxiv17}.

\emph{Summary and Discussion.}~~~In  this work we have shown that the
 observed
 gap-like behavior of the $B_{1g}$ Raman response in FeSe in a nematic metal is a direct consequence of
 the change of orbital composition of the pockets, which become nearly mono-orbital at $T \ll T_n$.
   The change in the orbital content of the pockets induces the
    angle-independent  component of the $B_{1g}$ Raman form-factor at the expense of the original
     $\cos{2\theta}$ component.
 The Raman intensity $\R$ contains only the non-symmetric $d$-wave part of the form-factor. When pockets become nearly mono-orbital,  the $d$-wave component of the form-factor gets strongly reduced
  at
   $\Omega \leq 2-3 \Delta_h$,
   and the $B_{1g}$  Raman intensity drops.

   To put our results in a broader context, we note that so far the orbital reconstruction of fermionic excitations in FeSe due to nematicity  has been
   proven only by means of polarized ARPES measurements~\cite{Watson_prb18,Liu_PRX2018}, which are sensitive to the surface. Our work shows that the Raman spectroscopy,
   which is a bulk probe, shows evidence of the same effect. Namely,  the suppression of the Raman response at $T$ below $T_n$
 necessarily implies a strong mixing between $d$-wave and $s$-wave channels, and we argue that
  the mixing
  is the consequence  of the change of the orbital composition of fermionic excitations.
   In this respect, our findings  are inconsistent with the scenario of  orbital-dependent spectral weights,  put forward in
      Refs.\ [\onlinecite{Hirschfeld2017,Davis2017,Qimiao_prb18, Hirschfeld_prb18,dai_natmat19}].   Within that scenario,  the orbital weights
        $u^2_\bk$ and $|v_\bk|^2$ would be further rescaled by $Z$. This rescaling would act against the reduction of the $yz$ spectral weight on the outer hole pocket and prevent this pocket from becoming mono-orbital.
         Then the Raman form-factor
           would
           retain its $d$-wave form, and the Raman response would remain largely  the same in the nematic phase, in disagreement with the data.

The reduction of $\R$ at $T < T_n$ due to orbital transmutation is quite generic and
should hold for any system undergoing a nematic transition. However, the strength of the transmutation and associated drop of $\R$ in the
   nematic phase    depends on details of the band structure. It would be highly desirable to analyze this scenario
     for other iron-based systems.

\begin{acknowledgments}
We are thankful to B. Andersen, A. Bohmer, G. Blumberg, M. Christensen, R. Fernandes, P. Hirschfeld, Y. Gallais, J. Kang, A. Klein, A. Kreisel, I. Paul,  A. Sacuto, and H. Yamase
for useful discussions.  This work has been supported by the Office
of Basic Energy Sciences, U.S. Department of Energy, under award
 DE-SC0014402, by the Italian MAECI under the Italian-India
collaborative  project  SUPERTOP-PGR04879, and by the Italian MIUR project PRIN 2017 n. .2017Z8TS5B.
AVC thanks for hospitality the Sapienza University of Rome, where this work was initiated.
\end{acknowledgments}

\bibliography{pnictides_nematic_new}

\begin{thebibliography}{51}
\bibitem [{com({\natexlab{a}})}]{comment1}%
  \BibitemOpen
  \href@noop {} {} ({\natexlab{a}}),\ \bibinfo {note} {by symmetry, there must
  exist a second hole band, constructed of $d_{xz}$ and $d_{yz}$
  orbitals.However it has been experimentally established that the inner pocket
  sinks below the Fermi level already in the tetragonal phase.}\BibitemShut
  {Stop}%
\bibitem [{com({\natexlab{b}})}]{comment2}%
  \BibitemOpen
  \href@noop {} {} ({\natexlab{b}}),\ \bibinfo {note} {the system may also
  develop $d$-wave nematic order on the outer d$_{xy}$ pocket (refs). Such
  order has not been unambigiously detected in ARPES studies and we will not
  consider it.}\BibitemShut {Stop}%
\end{thebibliography}

\newpage

\begin{widetext}
\begin{center}
{\bf {SUPPLEMENTARY INFORMATION}} \\
\end{center}

\section{BAND STRUCTURE AND ORBITAL WEIGHTS}
According to Eq.\ (2) in the main text the low-energy electronic structure of FeSe around the $\Gamma$ point can be described by the following approximated Hamiltonian $\hat H$ in the orbital space
\be
\lb{matrix}
\hat H = \begin{pmatrix} h_0(\bk)+h_3(\bk)+\Delta_h & h_1(\bk)-i\eta \\ h_1(\bk)+i\eta & h_0(\bk) -h_3(\bk)-\Delta_h \end{pmatrix},
\ee
where the $h_{i}$ components explicitly read
\be
\lb{param}
 h_0(\bk)= \epsilon_0- k^2/(2m), \hspace{1cm} h_1(\bk)= 2ck_xk_y, \hspace{1cm} h_3(\bk)= -b(k_x^2-k_y^2),
\ee
$\eta$ is the spin-orbit (SO) interaction and $\Delta_h$ the nematic order parameter at the $\Gamma$ point.
By a straightforward diagonalization of the Hamiltonian (\ref{matrix}), one recovers the eigenvalues
\be
\lb{bands}
E_{\pm,\bk}= h_0(\bk) \pm h(\bk) = h_0(\bk) \pm \sqrt{h_1(\bk)^2+(h_3(\bk)+\Delta_h)^2+\eta^2},
\ee
defining the band dispersion of the inner/outer hole pocket. In Table \ref{table1} we report the set of band parameters used to better reproduce the experimental measurements. We introduce $a\equiv 1/(2m)$ and set $b=c$, meaning that the pockets are assumed to be circular in the tetragonal phase. The nematic order parameter is assumed to be temperature-dependent below $T_n$ is a mean-field fashion. For practical purposes we use the approximated analytical expression $\Delta_h=\Delta_{h,0} (1-x^4/3)\sqrt{1-x^4}$, with $x=T/T_n$. The value of $\epsilon_0$ changes when the magnitude of the nematic order increases in such a way that the top of the outer pocket is kept approximately fixed at $\sim9,10$ meV. This is consistent with the analysis of Ref.\ \cite{Fanfarilloprb16}, considering that $\epsilon_0$ includes already the temperature-dependent isotropic self-energy correction due to spin fluctuations, responsible for the shrinking of the hole pockets both in the tetragonal and nematic phase. This effect, along with the splitting of the two pockets at  the zone center  due to SO interaction, is responsible for the fact that the inner hole pocket sinks below the Fermi level already in the tetragonal phase.

\begin{table}
\begin{tabular}{| c | c | c |}
\hline
$\Gamma$ pocket & $T \ge T_n$ (meV) & $T\ll T_n$ (meV) \\
\hline
$\epsilon_0$  & 0 & -9  \\
$a$ & 263 & 263\\
$b$ & 182 & 182\\
$\eta$ & 10 & 10\\
$\Delta_h$ & 0 & 15\\
\hline
\end{tabular}
\caption{Band parameters for the hole pockets at $\Gamma$ in the nematic and in the tetragonal phase.}
\label{table1}
\end{table}

The full Hamiltonian in the band basis reads $\sum_{\bk,i} E_\bk^i d_{i,\bk}^{\dag}d_{i,\bk}$, where $i=+/-$ denotes the outer/inner pocket, and the band operators can be written in terms of the orbital operators as
\be
\lb{transf}
\begin{aligned}
d_{-,\bk}&=u_\bk d_{xz,\bk} - v_\bk d_{yz,\bk}\\
d_{+,\bk}&= v_\bk^* d_{xz,\bk} + u_\bk d_{yz,\bk},
\end{aligned}
\ee
with orbital weights ($u_\bk,v_\bk$) explicitly given by
\be
\lb{bogol}
u_{\bk}= \frac{1}{\sqrt{2}} \left( 1+ \frac{h_3(\bk)-\Delta_h}{h(\bk)} \right)^{1/2}, \hspace{1cm} v_{\bk}= \frac{h_3(\bk)-\Delta_h-h(\bk)}{h_1(\bk)+i\eta}u_{\bk}.
\ee
Notice that $u_{\bk}^2+|v_{\bk}|^2=1$ as usual, so that $|v_{\bk}|^2=1/2\left[1-(h_{3}(k)-\Delta_h)/h(k)\right]$. For the sake of compactness, we thus define $u_\bk^2\equiv \cos^2{\bar \theta_\bk}$ and $|v_\bk|^2 \equiv \sin^2{\bar \theta_\bk}$. As a fingerprint of $C_4$ symmetry, in the tetragonal phase ($\Delta_h=0$) and in absence of SO coupling ($\eta=0$), one immediately finds that $u_{\bk}^2 = \cos^2 \theta$ and $|v_{\bk}|^2=\sin^2 \theta$, being $\theta$ the angle along the pockets, such that  $k_x=k\cos{\theta}$, $k_y=k\sin{\theta}$. On the contrary, when $\lambda= \Delta_h/(bk^2) \gg 1$ deep inside the nematic phase, one notices that $u_\bk^2 \simeq 0$ and $|v_\bk|^2 \simeq 1$ regardless of $\theta$, meaning that in this limit the orbital character becomes almost entirely $d_{xz}$ in the outer pocket and $d_{yz}$ in the inner one (see Fig.\ \ref{uv}a). When SO interaction is included, the two orbital characters get mixed, the transition to the mono-orbital configuration in the nematic phase is smoothened but the effect is still robust, as long as $\lambda$ is large enough (see Fig.\ \ref{uv}b-c).
\begin{figure}[h!]
\centering
\includegraphics[scale=0.4]{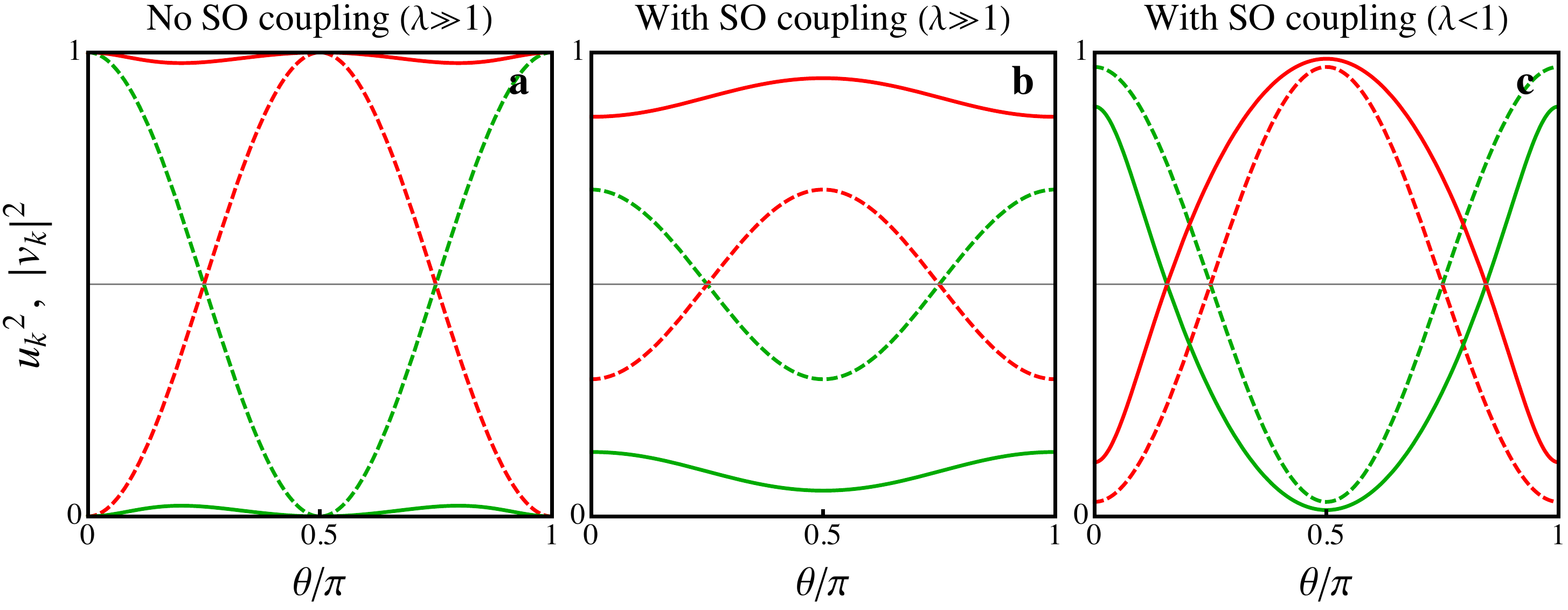}
\caption{Orbital content of the $\Gamma$ pocket as a function of $\theta$ in the tetragonal (dashed lines) and in the nematic phase (solid lines). Red lines refer to $|v_k|^2$, green lines to $u_k^2$.}
\label{uv}
\end{figure}

\section{COMPUTING THE FULL $\bm{B_{1g}}$ RAMAN RESPONSE}

As it is well known\cite{devereauxreview}, the Raman response can be related to a density-like correlation function, where the density operator is weighted with a momentum-dependent factor which accounts for the incoming/outgoing light polarization. Starting from the orbital model \pref{matrix},  the $B_{1g}$ Raman vertex  can then be defined as $\Gamma_{B_{1g}} \equiv [\partial H/\partial k^2_x - \partial H/\partial k^2_y ]$. By using the explicit expressions for the $h_i(\bk)$ functions one then finds that $\hat \Gamma_{B_{1g}}\propto \hat \tau_3$ in the orbital basis. Using then Eq.\ (\ref{transf}), the $B_{1g}$ Raman density can be rewritten in the band basis as
\be
\label{R_de}
d_{xz,\bk}^{\dag}d_{xz,\bk}-d_{yz,\bk}^{\dag}d_{yz,\bk} = (u_\bk^2-|v_\bk|^2)(d_{-,\bk}^{\dag}d_{-,\bk}-d_{+,\bk}^{\dag}d_{+,\bk})+2u_\bk v_\bk d_{-,\bk}^{\dag}d_{+,\bk}+2u_\bk v_\bk^*d_{+,\bk}^{\dag}d_{-,\bk},
\ee
showing that the Raman density operator in the ${B_{1g}}$ channel has  both an intraband and an interband component. In particular, one can define
\be
\label{Gamma}
\Gamma_{B_{1g}}^{intra}(\bk,\theta)  \propto u_\bk^2-|v_\bk|^2 = \frac{h_3(\bk)-\Delta_h}{h(\bk)} \equiv \cos{2\bar \theta_\bk}, \hspace{1cm}
\Gamma_{B_{1g}}^{inter}(\bk,\theta)  \propto  -2u_\bk v_\bk^* = -\frac{h_1(\bk)+i\eta}{h(\bk)},
\ee
so that $|\Gamma_{B_{1g}}^{inter}|^2 \propto 1 - \cos^2{2\bar \theta_{\bk}} \equiv \sin^2{2\bar \theta_{\bk}}$. The full $B_{1g}$ Raman response is associated with the imaginary part of the density-density correlation function with two Raman vertices, evaluated in the long-wavelength limit ($\bq=0$), i.e.
\be
\label{full}
\chi_{B_{1g}}(\bq=0,\Omega)=\chi_{B_{1g}}^{intra}(\Omega)+\chi_{B_{1g}}^{inter}(\Omega),
\ee
where
\bea
\lb{intra}
\chi_{B_{1g}}^{intra}(\Omega) & \propto & i \int \frac{d^2k}{(2\pi)^2} \cos^2{2\bar \theta_{\bk}} \int \frac{d\omega}{2\pi} \left[ \Gr_+(\bk,\omega+\Omega/2) \Gr_+(\bk,\omega-\Omega/2) + \Gr_-(\bk,\omega+\Omega/2) \Gr_-(\bk,\omega-\Omega/2) \right] \label{intra}\\
\lb{inter}
\chi_{B_{1g}}^{inter}(\Omega) & \propto & i \int \frac{d^2k}{(2\pi)^2} \sin^2{2\bar \theta_{\bk}} \int \frac{d\omega}{2\pi} \left[ \Gr_+(\bk,\omega+\Omega/2) \Gr_-(\bk,\omega-\Omega/2) + \Gr_-(\bk,\omega+\Omega/2) \Gr_+(\bk,\omega-\Omega/2) \right], \label{inter}
\eea
where $\Gr_{\pm}(\bk,\omega) \equiv [\omega - E_{\pm,\bk} + i\gamma \mbox{sng}(\omega)]^{-1}$ is the $T=0$ electronic Green's function and $\gamma >0$ is the scattering rate due to impurities. The tendency towards nematic instability can be explained by adding a finite attraction $U$ acting in the intraband sector only. As a consequence, the full $B_{1g}$ Raman response reads:
\bea
R_{B_{1g}}^{intra} & \propto & \mbox{Im}\left[\frac{\chi^{intra}_{B_{1g}}}{(1-U\chi^{inter}_{B_{1g}})}\right],\\
R_{B_{1g}}^{inter} & \propto & \mbox{Im}{[\chi^{inter}_{B_{1g}}]}.
\eea
The results for $R_{B_{1g}}$ are shown in Fig.\ 2 of the main manuscript. As already discussed there, one finds that at small frequencies the interband component is not relevant and the intraband contribution associated with particle-hole excitations in the inner pocket is strongly sub-leading compered to the outer pocket one. Therefore, let us focus on the (outer pocket) intraband component only. The integrals over frequency in Eq.s\ \pref{intra}-\pref{inter} can be solved analytically in a straightforward way, while the integration in the momentum space is performed numerically. More specifically, for the intraband part we have
\be
\label{intra2}
\chi_{B_{1g}}^{intra}(\Omega) \propto  -\frac{\gamma}{\Omega (\Omega+2i\gamma)} \int \frac{d\theta}{2\pi} \int \frac{kdk}{2\pi} \cos^2{2\bar \theta_{\bk}} \log{\left[\frac{E_{+,\bk}^2-(\Omega+i\gamma)^2}{E_{+,\bk}^2+\gamma^2}\right]}.
\ee
In the tetragonal phase ($\Delta_h=0$) and in absence of SO interaction ($\eta=0$), the band dispersion is parabolic and angle independent, i.e.\ $E_{+,\bk}=\epsilon_0-k^2/(2m)+bk^2=\epsilon_0-k^2/(2m^*)$. In addition since $\cos{2\bar \theta_{\bk}} \equiv u_{\bk}^2-|v_{\bk}|^2 = \cos^2{\theta}-\sin^2{\theta}=\cos{2\theta}$ the intraband Raman tensor depends only on the angle. In this situation the angle and momentum integration in Eq.\ \pref{intra2} are decoupled, and the Raman polarization factor gives simply a constant pre-factor when integrated over $\theta$. The integration over momenta can also be performed analitically, and the final results depends in general on $\epsilon_0/\Omega$ and $\epsilon_0/\gamma$. In the limit when both ratios are much larger than 1, so that finite-band effects are negligible and the Fermi energy is the largest energy scale in the problem, one recovers the well-known analytical result  mentioned in the main text, i.e.
\be
\lb{intra3}
\chi_{B_{1g}}^{intra}(\Omega) \sim  \frac{2 i \gamma}{\Omega+2 i \gamma}.
\ee
This can be seen by rewriting the first term in Eq.\ (\ref{intra})  (up to constant prefactors) as
\be
\lb{RARRAA}
\begin{aligned}
\chi_{B_{1g}}^{intra}(\Omega) & \propto  i \int_{-\epsilon_0}^{+\infty} d E_{+} \Bigg[ \int_{-\infty}^{-\omega/2} \frac{d\omega}{2\pi} \Gr_{+}^{A}(E_{+},\omega+\Omega/2) \Gr_{+}^{A}(E_{+},\omega-\Omega/2) +  \int_{-\Omega/2}^{+\Omega/2} \frac{d\omega}{2\pi} \Gr_{+}^{R}(E_{+},\omega+\Omega/2) \Gr_{+}^{A}(E_{+},\omega-\Omega/2) +\\
 & +\int_{\Omega/2}^{+\infty} \frac{d\omega}{2\pi} \Gr_{+}^{R}(E_{+},\omega+\Omega/2) \Gr_{+}^{R}(E_{+},\omega-\Omega/2) \Bigg] = \chi_{B_{1g}}^{AA}+\chi_{B_{1g}}^{RA}+\chi_{B_{1g}}^{RR},
\end{aligned}
\ee
where $\Gr^A_{+}$/$\Gr^B_{+}$ denote the advanced/retarded Green's functions respectively (i.e. $\mbox{sng}(\omega \pm \Omega/2) = -/+$). By performing the integration over frequency fist, explicit analytical calculations lead to
\bea
\lb{RA}
\chi_{B_{1g}}^{RA}(\Omega) & =& i \int_{-\epsilon_0}^{+\infty} dE_{+} \frac{1}{2\pi(\Omega+2i\gamma)}\left[ \log{\frac{-E_{+}-i\gamma}{\Omega-E_{+}+i\gamma}}+\log{\frac{-E_{+}+i\gamma}{-\Omega-E_{+}-i\gamma}} \right] \stackrel{\epsilon_0 \to \infty}{=} -\frac{\Omega}{\Omega+2i\gamma}
\\
\lb{RR+AA}
\chi_{B_{1g}}^{AA}+\chi_{B_{1g}}^{RR} & = & i \int_{-\epsilon_0}^{+\infty} dE_{+} \frac{1}{2\pi \Omega} \left[ \log{\frac{E_{+}+\Omega+i\gamma}{E_{+}+i\gamma}}+\log{\frac{E_{+}-\Omega-i\gamma}{E_{+}-i\gamma}} \right] \stackrel{\epsilon_0 \to \infty}{=} 1.
\eea
In the case of FeSe the Fermi level is rather small, and in addition the presence of SO interaction leads to deviation from a pure parabolic behvaior already in the tetragonal phase. Nonetheless, the direct comparison between the numerical solution of Eq.\ (\ref{intra2}) at $T>T_n$ and Eq.\ \pref{intra3} shows that, despite some quantitative deviation, the qualitative behavior of the Raman bubble pretty much follows the analytical expression. For this reason, while computing the vertex corrections we will rely on the infinite-bandwidth limit for $\chi_{B_{1g}}^{intra}$, that allows one for an analytical solution of the self-consistent vertex equation.

\begin{figure}[h!]
\centering
\includegraphics[scale=0.55]{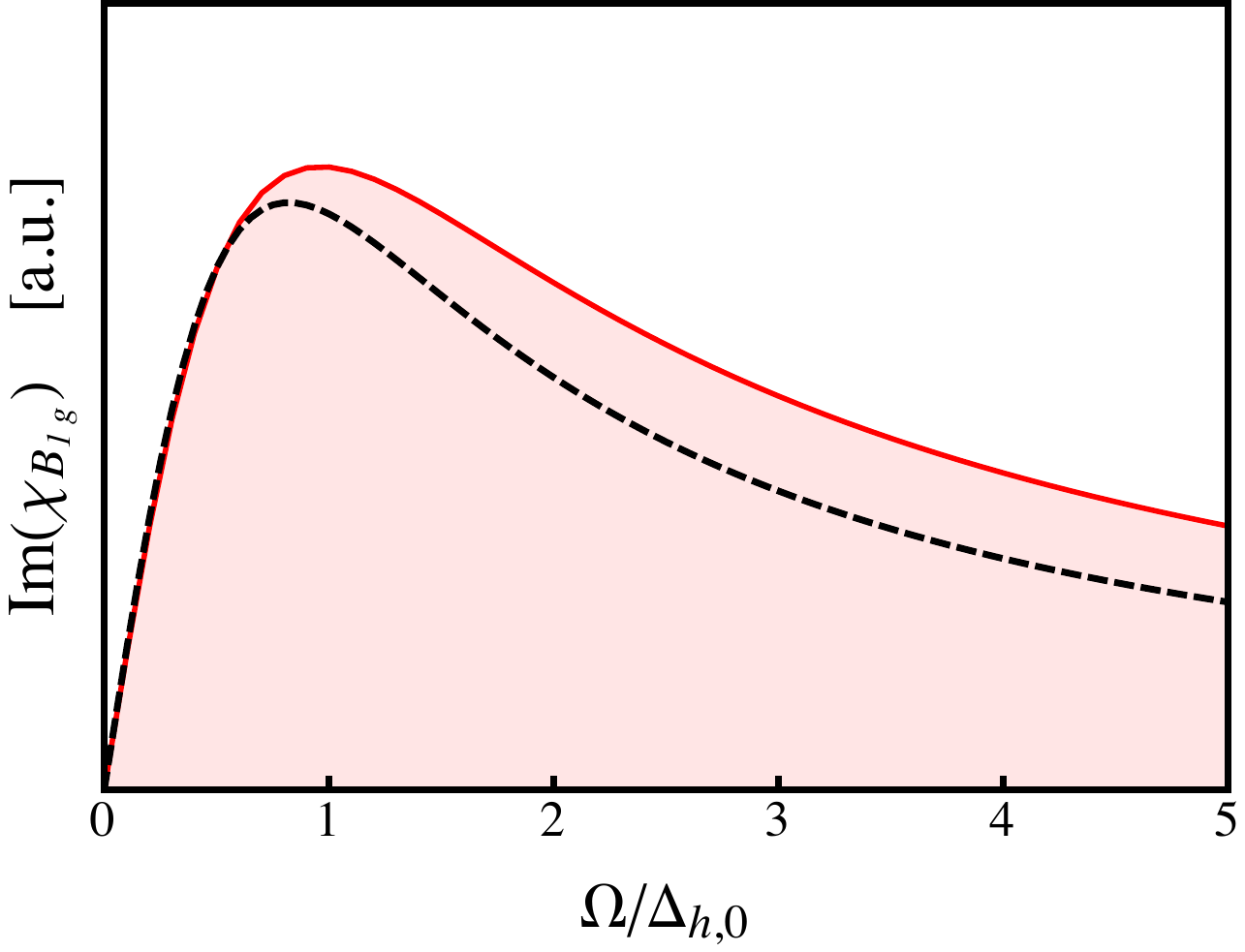}
\caption{Comparison between the analytical result (\ref{intra3}) (black, dashed line) and the numerical solution of Eq.\ (\ref{intra2}) (red plain line), obtained by using the band parameters reported in Table S1 at $T\ge T_n$ and by fixing the electron broadening at $\gamma \simeq 6$ meV. }
\label{analytics}
\end{figure}

\section{VERTEX CORRECTIONS IN THE LADDER APPROXIMATION}

To understand the role of vertex corrections, let us focus now on the density-density response $\chi(\bq=0, \omega)$ for a  single-band model.  In a perfect metal, the simultaneous conservation of energy and momentum implies that intraband particle-hole excitations at zero external momentum transfer ($\bq=0$) are not permitted. However, in the presence of impurities and/or interactions the electronic energy levels get broader in a range of order $\gamma$, allowing for a nonzero $\chi(\omega)$  at energies lower than $\sim 2\gamma$, see Eq.\ \pref{intra3}. As discussed in the previous Section, this result has been obtained by including self-energy effects in the electronic Green's functions $\Gr$. Indeed, in the presence of  impurities the scattering rate $\gamma$ introduced in the Green's function is given by the imaginary part of the self-energy at zero frequency computed in the Born approximation\cite{mahan_book}, so that it is explicitly given by
\be
\label{broad}
\gamma \equiv \pi n_i V_0^2 N_F,
\ee
where $n_i$ is the concentration of impurities, $N_F$ the electron density at the Fermi level and $V_0$ the scattering potential, assumed to be momentum and angular-independent for simplicity. As a consequence, including self-energy corrections one obtains for the density-density correlation function $\chi(\omega)$  the same result as Eq.\ \pref{intra3} above, i.e.
\be
\lb{SE}
\chi^{se}(\Omega) = \chi^{se}_{RR}+\chi^{se}_{AA}+\chi^{se}_{RA}= N_F \left(1- \frac{\Omega}{\Omega+2i\gamma}  \right),
\ee
where we used the same decompositon in R/A components introduced in Eq.s \pref{RA}-\pref{RR+AA} above.
However, it is evident that Eq.\ \pref{SE} cannot be the final response, since charge conservation implies that $\chi(\omega)=0$ at any finite frequency. The reason is very simple: the density-density correlation function at $\bq=0$ at finite frequency represents the response to a uniform potential. Due to charge conservation, changing
the charge density in one place can only be done by redistributing
it, but this cannot be achieved with a uniform potential. As a consequence $\chi(\bq=0,\omega)=0$ must vanish identically.
To recover this result, one has to include, along with the self-energy corrections \pref{SE}, the vertex corrections due to disorder, by summing up the series of particle-hole bubbles with increasing number of scattering events linking the Green's functions at opposite sides of the bubble, i.e.\ the so-called \emph{ladder diagrams}, whose resummation gives
\be
\lb{eqver}
\chi^v(\Omega)=i\int \frac{d^2k}{(2\pi)^2} \frac{d\omega}{2\pi}  \Gr(\bk,\omega,\omega+\Omega)\Gr(\bk,\omega) n_i V_0^2 \frac{1}{1-n_i V_0^2 \int \frac{d^2k'}{(2\pi)^2}   \Gr(\bk',\omega,\omega+\Omega)\Gr(\bk',\omega)}.
\ee
The full density response is then given by
\be
\lb{full}
\chi(\Omega)=\chi^{se}(\Omega) + \chi^{v}(\Omega).
\ee
As shown e.g.\ in Ref.\ \cite{aleiner}, the relevant vertex corrections act on the $RA$ component of the fermionic bubble. In particular, since in the infinite-bandwidth approximation
\be
\int \frac{d^2k}{(2\pi)^2}  \Gr^R(\bk,\omega,\omega+\Omega)\Gr^A(\bk,\omega)=\frac{2\pi i N_F}{\O+2i\gamma},
\ee
and the range of frequency integration for $\chi^v_{RA}$ in Eq.\ \pref{eqver} is between $-\O/2$ and $\O/2$, we get that
\be
\lb{vsol}
\chi^v=N_F\frac{-2i\gamma}{\O+2i\gamma}.
\ee
By adding this term to Eq.\ \pref{SE} we see that $\chi^{se}+\chi^v=0$, restoring gauge invariance as expected.

Let us see now how this result is relevant for our problem. In the spirit of Eq.\ (\ref{intra3}), for the sake of analytical calculations one can  model the intraband part of the $B_{1g}$ Raman response as
\be
\lb{SEr}
\chi_{B_{1g}}^{se}(\Omega) = \langle \Gamma_{B_{1g}}^2(\Omega) \rangle N_F \left(1- \frac{\Omega}{\Omega+2i\gamma}  \right),
\ee
where the $se$ superscript is used to remember that this is the result obtained adding only self-energy effects induced by the presence of impurities. The dependence on the Raman polarization has been factorized out, and included in the prefactor
\be
\label{raman_vertexSQ}
\langle \Gamma_{B_{1g}}^2(\Omega) \rangle \equiv \int \frac{d\theta k dk}{(2\pi)^2} \frac{\delta(\Omega-E_{\bk}(\theta))}{N(\Omega)} (\cos{2\bar \theta_{\bk}})^2,
\ee
with $N(\Omega)= \int \frac{d\theta kdk}{(2\pi)^2} \delta(\Omega - E_{\bk}(\theta))$. As we discussed in the main text and in Sec.\ S1 above, the orbital composition of the pocket in the nematic phase, and consequently the angular dependence of $\Gamma_{B_{1g}}$, is strongly influenced by the ratio $\lambda = \Delta_h/(bk^2)$: when $\lambda <1$ the two orbitals equally contribute to band excitations, i.e.\ $u_{\bk}^2 \sim \cos^2{\theta}$, $|v_{\bk}|^2 \sim \sin^2{\theta}$ and the Raman vertex is essentially $d-$wave symmetric ($\cos{2\bar \theta_{\bk}} \propto \cos{2\theta}$), like in the tetragonal phase; when instead $\lambda \gg 1$, the pocket is almost entirely $d_{xz}$, $u_{\bk}^2 \simeq 0$, $|v_{\bk}|^2 \simeq 1$ and the Raman vertex develops an angle-independent ($s-$wave) component, i.e.\ $\cos{2\bar \theta_{\bk}} \approx -1$. Therefore, one can conveniently write $\cos{2\bar \theta_{\bk}} = \Gamma_{s,\bk}+\Gamma_{d,\bk} \cos{2\theta}$, so that $\langle \Gamma_{B_{1g}}^2 \rangle$ can be explicitly decomposed in the $d-$wave and the $s-$wave component
\be
\lb{dec}
\langle \Gamma_{B_{1g}}^2 \rangle = \Gamma_s^2+\Gamma_d^2.
\ee
Obviously, mixed terms $ \Gamma_s \Gamma_d = \langle\Gamma_{s,\bk} \Gamma_{d,\bk} \cos{2\theta} \rangle$ are always zero regardless the specific $k$-dependence of the Raman vertex, since $\int d\theta \cos{2\theta}=0$. In the tetragonal phase $\langle \Gamma_{B_{1g}}^2 \rangle = \Gamma_d^2$, while in the nematic phase $\langle \Gamma_{B_{1g}}^2 \rangle \approx \Gamma_s^2$ if $\lambda \gg 1$ and $\langle \Gamma_{B_{1g}}^2 \rangle \approx \Gamma_d^2$ when $\lambda < 1$.

Once established the general decomposition \pref{dec} we can compute again the vertex corrections due to disorder in the same spirit of Eq.\ \pref{eqver} before. However, in this case we have to consider the presence of a Raman $\Gamma_{B_{1g}}$ vertex on each side of the bubble, see Fig.\ \ref{VCOR}. Then Eq.\ \pref{eqver} should be written in this case:
\be
\lb{eqver2}
\chi^v(\Omega)=i\int \frac{d^2k}{(2\pi)^2} \frac{d\omega}{2\pi} \Gamma_{B_{1g}}(\bk) \Gr(\bk,\omega,\omega+\Omega)\Gr(\bk,\omega) n_i V_0^2 \frac{1}{1-n_i V_0^2 \int \frac{d^2k'}{(2\pi)^2}  \Gamma_{B_{1g}}(\bk')  \Gr(\bk',\omega,\omega+\Omega)\Gr(\bk',\omega)}.
\ee
By transforming again the momentum integration in the angular average times the energy integration one easily sees that only the $\Gamma_s$ term survives in the vertex part, since $\langle \Gamma_d\rangle=0$. As a consequence, since $\langle \Gamma_s\rangle^2=\langle \Gamma_s^2\rangle$ the vertex corrections read in this case
\be
\lb{eqverr}
\chi^v=-N_F \Gamma_s^2 \frac{2i\gamma}{\O+2i\gamma},
\ee
 which means that vertex corrections  are not relevant  when the Raman vertex is purely $d-$wave symmetric, i.e.\ in the tetragonal phase and in the nematic phase when $\lambda$ is small. By summing Eq.\ \pref{SEr} and \pref{eqverr} one obtains Eq.\ (6) of the main manuscript, i.e.
 \be
 \lb{full}
 \chi_{B_{1g}}(\Omega)= N_F \Gamma_d^2  \frac{2i\gamma}{\Omega+2i\gamma}. 
\ee
To obtain the $\Gamma_s$ component we have to compute the angular average of the Raman vertex, as detailed in Eq.\ (7) of the main manuscript:
\be
\lb{eqgammas}
\Gamma_s\equiv \langle \Gamma_{B_{1g}}(\Omega) \rangle =\int \frac{d\theta k dk}{(2\pi)^2} \frac{\delta(\Omega-E_{\bk}(\theta))}{N(\Omega)} \cos{2\bar \theta_{\bk}}.
\ee
The $\Gamma^2_d$ component is instead obtained as the difference between the average of the Raman vertex squared and its mean value \pref{eqgammas}. As a consequence, the full result \pref{full} can also be written as:
\be
\label{final}
\chi_{B_{1g}}(\Omega)= N_F \frac{2i\gamma}{\Omega+2i\gamma} \left( \langle \Gamma_{B_{1g}}^2(\Omega) \rangle - \langle \Gamma_{B_{1g}}(\Omega) \rangle^2  \right).
\ee
\begin{figure}[h!]
\centering
\includegraphics[scale=0.55]{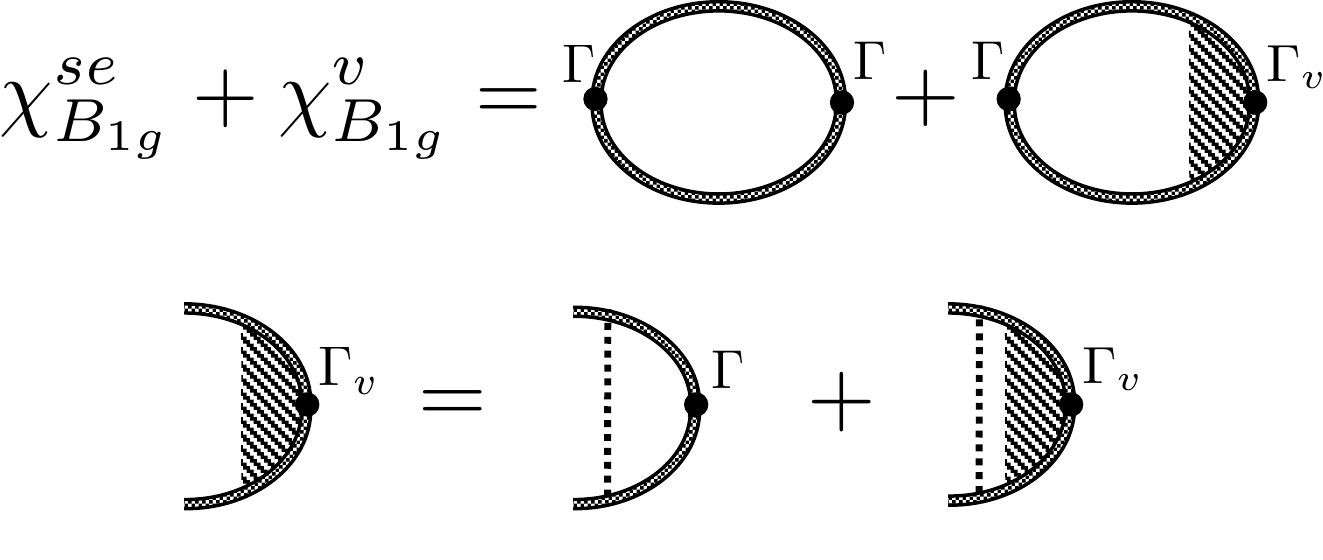}
\caption{Self-energy and vertex-correction contributions to the Raman response. On the bottom line we show the diagrammatic representation of the self-consistent equation satisfied by the dressed vertex $\Gamma_v$.}
\label{VCOR}
\end{figure}

\end{widetext}

\end{document}